\documentstyle[11pt,epsfig]{article}
\newcommand{\ee}{\mbox{$e^+e^-$}}
\newcommand{\beq}{\begin{equation}}
\newcommand{\eeq}{\end{equation}}
\newcommand{\bfg}{\begin{figure}}
\newcommand{\efg}{\end{figure}}
\newcommand{\Eg}{\mbox{$E_\gamma$}}
\newcommand{\dn}{\mbox{$\Re (n_\parallel-n_\perp)$}}
\title{\bf Measurement of the Longitudinal Polarization of the
             HERA Electron Beam Using Crystals and\\ the ZEUS Luminosity
           Monitor\footnote{Contribution to the Workshop on Channeling
           and Other Coherent Crystal Effects at Relativistic
           Energies, \AA rhus, Denmark, July 10--14, 1995}\\}
\author{{\sf Krzysztof Piotrzkowski}\\ \small \it Institute of Nuclear
Physics, PL--30055 Cracow, Poland\\
\small \it and\\ \small \it DESY, D--22607 Hamburg, Germany}
\begin{document}
\maketitle
\begin{abstract}
A measurement of the longitudinal polarization of the electron beam
at HERA utilizing coherent interactions of high energy photons in
crystals is described. Modification of existing facilities would allow
an independent polarization measurement and a verification of
birefringence phenomena in crystals for 20--30~GeV photons. Relevant
experimental issues and systematic uncertainties are also presented.
\end{abstract}
\section{Introduction}
High longitudinal polarization ($>60$\%) of the electron beam has been
routinely achieved at HERA since the installation of so-called
spin rotators which convert the naturally acquired transverse polarization
to the experimentally required longitudinal polarization \cite{pol1}.
Longitudinal polarization is currently available only at the interaction
point of the fixed-target experiment HERMES; spin rotators are planned for
installation
at the collider experiments, ZEUS and H1. The success of the physics program
with polarized electrons (and possibly polarized protons) at HERA relies
heavily on the precision of the polarization measurement.

Measurement of the polarization of the electron beam at HERA is
currently performed by Compton-scattering laser photons off the electron
beam and measuring the spatial asymmetry of the scattered photons \cite{pol2}.
The estimated accuracy of this method is about 2\%. An independent
polarization measurement is of interest as it would allow for better control of
systematic errors and cross-calibration of the polarimeters\footnote{Recently,
a measurement of asymmetries in large angle ep bremsstrahlung
was proposed as an interesting method of polarization monitoring at
colliders \cite{arbuzov}.}.

In 1988 the measurement of the electron beam polarization using
bremsstrahlung attenuation in crystals was proposed for use at the
LEP collider \cite{lep}. We developed a similar method for use at HERA
utilizing an existing detector of high energy bremsstrahlung which is
presently used for the luminosity measurement in the ZEUS experiment
\cite{lumi1}.
\section{Principles of the Method}
It is aimed to achieve at least 2\% accuracy of the measurement
of the electron beam longitudinal polarization. The measurement
would be done in three steps:
\begin{enumerate}
\item calculation of the correlation between the electron longitudinal
polarization and the circular polarization of high energy
bremsstrahlung (including detector effects);
\item conversion of the photon polarization from circular
to linear (or at least to strongly elliptic);
\item measurement of the linear polarization of the photon.
\end{enumerate}
The first point is well understood and theoretical relations are available;
the second step has been demonstrated theoretically but has never been realized
experimentally, and it is very interesting on its own; the third step has
been demonstrated experimentally confirming the theoretical predictions,
therefore should not pose many problems.

For a longitudinally polarized electron beam, bremsstrahlung
photons with energies close to the incident electron energy
are circularly polarized to the same degree as the electron beam.
The bremsstrahlung polarization can be deduced for an initial electron
longitudinal polarization, $h_e^L$, and bremsstrahlung photon circular
polarization, $h_\gamma^C$, from the following differential
cross-section \cite{caffo}:
\beq
 \frac{\rm d\sigma}{\rm dy}=2\alpha
 r_0^2\left(\frac{1}{y}\left[\frac{4}{3}(1-y)+y^2\right]
 +h_\gamma^Ch_e^L\frac{1}{3}(4-y)\right)
  \left(\ln[4E_eE_p(1-y)/m_eM_py]-
  \frac{1}{2}\right)\, ,
\label{xsec}
\eeq
where y is the fraction of the initial electron energy, $E_e$, carried by
the bremsstrahlung photon, $E_p$ is proton beam energy, $M_p$ and
$m_e$ are the proton and electron masses, $\alpha$ is the fine structure
constant and $r_0$ is the classical electron radius, and it is assumed that
the polarization of the scattered electron is not measured. From Eq.
\ref{xsec} one can derive the following relation between the electron beam
longitudinal polarization\footnote{Beam polarization is defined as
$\frac{N_+-N_-}{N_++N_-}$, where $N_+, N_-$ are numbers of
particles with `+' and `--' polarization, respectively.}, $P_e^L$, and the
bremsstrahlung circular polarization, $P_\gamma^C$:
\beq
 P_\gamma^C=P_e^L\frac{y(4-y)}{4(1-y)+3y^2}\, .
\eeq
It follows that at y=1 $P_\gamma^C=P_e^L$, and that $P_\gamma^C$ is
near $P_e^L$ if y is close to 1, e.g. if
y=0.9 then $P_\gamma^C=0.986P_e^L$.

In 1962 Cabibbo et al. proposed the use of crystals for producing and
analyzing the linear polarization of high energy photons
\cite{cabbibo1,cabbibo3}.
Photons which impinge on a crystal under small angles with respect to the
crystallographic axes (or planes) undergo a coherent interactions with strings
of atoms residing in lattice sites when the direction of the recoil momentum,
q,
coincides with one of the crystal axes and its value is related to the
interplanar
distance a, $|q|=h/a$. For linearly polarized photons the cross-section for
\ee\ pair creation, and hence the photon attenuation in a crystal, depends
significantly on the angle between the plane of the photon polarization and
the crystal symmetry plane. A crystal can be used as a polarizer, since an
initially unpolarized photon beam becomes linearly polarized after traversing
the thick crystal, as shown experimentally at Cornell and SLAC
\cite{berger,eisele}. The crystal can also be used as an analyzer by measuring
the attenuation of the photon beam as a function of the crystal orientation.
One infers the linear polarization of the incident photon beam from the
variation of the photon attenuation for different crystal positions, e.g. if
the
incident photons have a linear polarization $\vec{P}$ and hit the crystal under
a small angle $\theta$ with respect to the (110) axis, then the attenuation of
the photon beam changes with angle $\phi$ between the polarization and
the (001) planes as follows \cite{cabbibo1}:
\beq
\frac{I(x)}{I(0)}=\left.\frac{I(x)}{I(0)}\right|_{\vec{P}=0}[1+|\vec{P}|A(x)cos2\phi],
\label{aeq}
\eeq
where I(0) is the initial intensity and I(x) is the intensity of the
photon beam after traversing the crystal thickness x, A is the crystal
analyzing power and is equal to the linear polarization of an initially
unpolarized photon beam after traversing the crystal. The crystal should be
rotated around the photon beam axis in such a way that the angle $\theta$ is
fixed.

However, bremsstrahlung is circularly polarized therefore to use
such crystal analyzers one needs to convert the polarization
of high energy photons from circular to linear polarization. A $\lambda/4$
polarization converter for high energy photons was proposed by
Cabibbo et al. \cite{cabbibo2} who discovered (and utilized) the property
of birefringence of crystals, i.e. the non-zero difference of the refractive
indices, \dn, for parallel and perpendicular polarizations (with respect to
the crystallographic planes or axes) of high energy photons.
The appropriate $\lambda/4$ converter thickness, d, can be obtained from
the following formula:
\beq
\dn\Eg d=\pi/2\, ,
\eeq
where $\Eg=yE_e$ is the photon energy. The converter is however
also a polarizer, therefore the linear polarization behind it,
$P_\gamma^L$, is higher than the initial photon circular polarization
$P_\gamma^C$ \cite{clarke}:
\beq
P_\gamma^L=\sqrt{A_c^2(d)+[1-A_c^2(d)](P_\gamma^C)^2}\, ,
\label{ceq}
\eeq
where $A_c$ is the converter analyzing power.

The other aim of the proposed experiment is to verify
experimentally the exciting theoretical prediction of crystal
birefringence for high energy photons. To show, in particular, that not
only visible light but also many-GeV photons can travel at a
speed smaller than the speed of light in vacuum! In fact, only when
this effect has been demonstrated can the proposed method of
polarization measurement be implemented.
\section{Experimental Set-Up}
A polarization measurement using crystals could be performed utilizing
existing detectors of the ZEUS luminosity monitor. A detailed description of
the luminosity detectors can be found elsewhere \cite{lumi1,lumi2}.  The
`branch'
used for the bremsstrahlung photon detection is briefly described here.
The photon detector is located about 100~m from the interaction point (IP),
upstream of the proton beam (see Fig. \ref{det}). It consists
of a $1.5$~mm thick circular (100~mm in diameter) copper-beryllium exit
window in the HERA vacuum chamber at 92~m from the IP, a 12~m long
vacuum pipe, a $2.5X_0$ carbon-lead absorber ($1X_0\equiv$ one
radiation length) shielding against synchrotron radiation, and a
$22~X_0$ lead-scintillator `sandwich' calorimeter at about 107~m
from the IP. The calorimeter sampling step is $1~X_0$ with the first
scintillator plate (`presampler') being 4 times thicker than the nominal
0.26~mm thickness, to compensate for the energy loss in the absorber
($3.7 X_0$ in total).
The signal from the calorimeter scintillator plates is read-out by
two (up and down) wavelength shifter plates connected by light
guides to two photomultiplier tubes. The calorimeter is equipped with
a position detector with separate readout at a depth of $3X_0$.
\bfg
\includegraphics{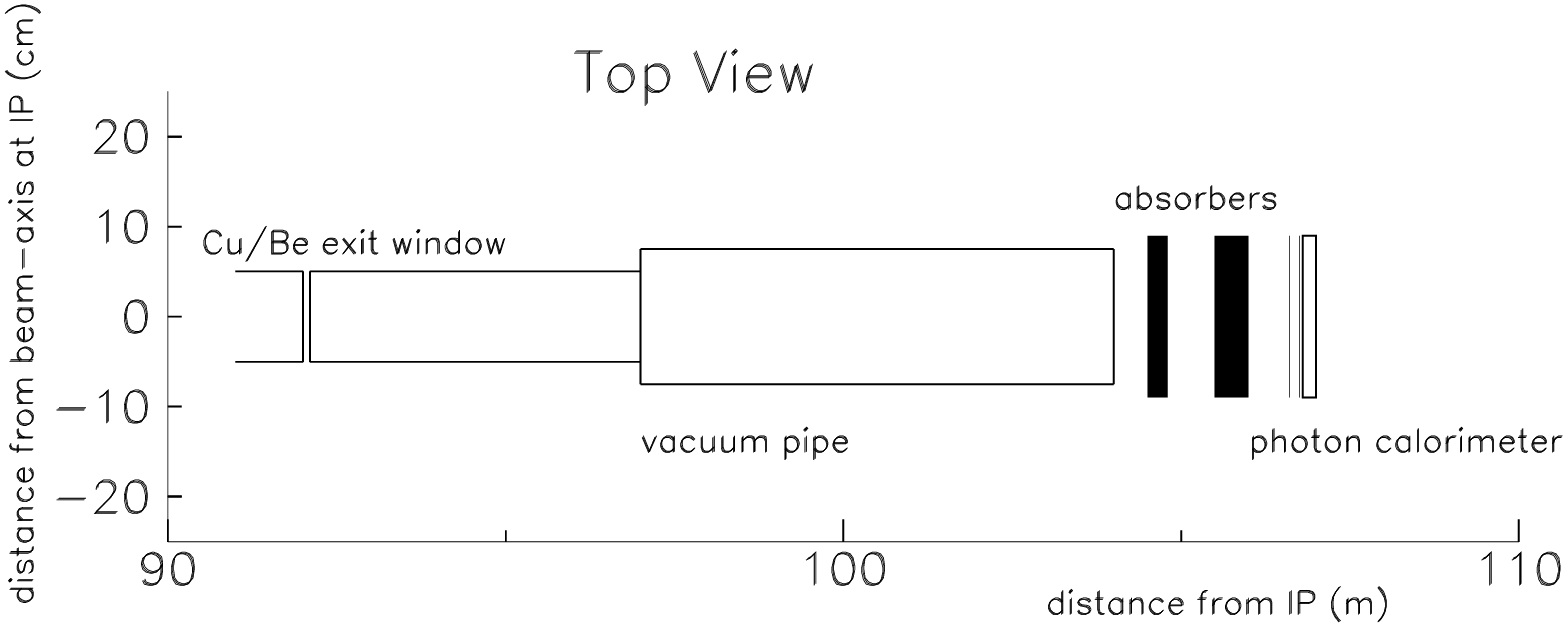}
\begin{picture}(140,55)(0,0)
\end{picture}
\caption{\sf Layout of the photon detector in the ZEUS luminosity
monitor.}
\label{det}
\efg

Here, we propose to change the experimental setup of the ZEUS
luminosity monitor in order to accommodate a simultaneous polarization
measurement. The modification requires replacement of the
carbon filter and the first 1~$X_0$ lead plate with two thick crystals
serving as the polarization converter and analyzer. Additionally, for
the detection of the photon interaction in the crystals (needed in the
photon attenuation measurement) a detector sensitive to charged particles
has to be placed behind the analyzer, see Fig. \ref{setup}.
\bfg
\epsfig{file=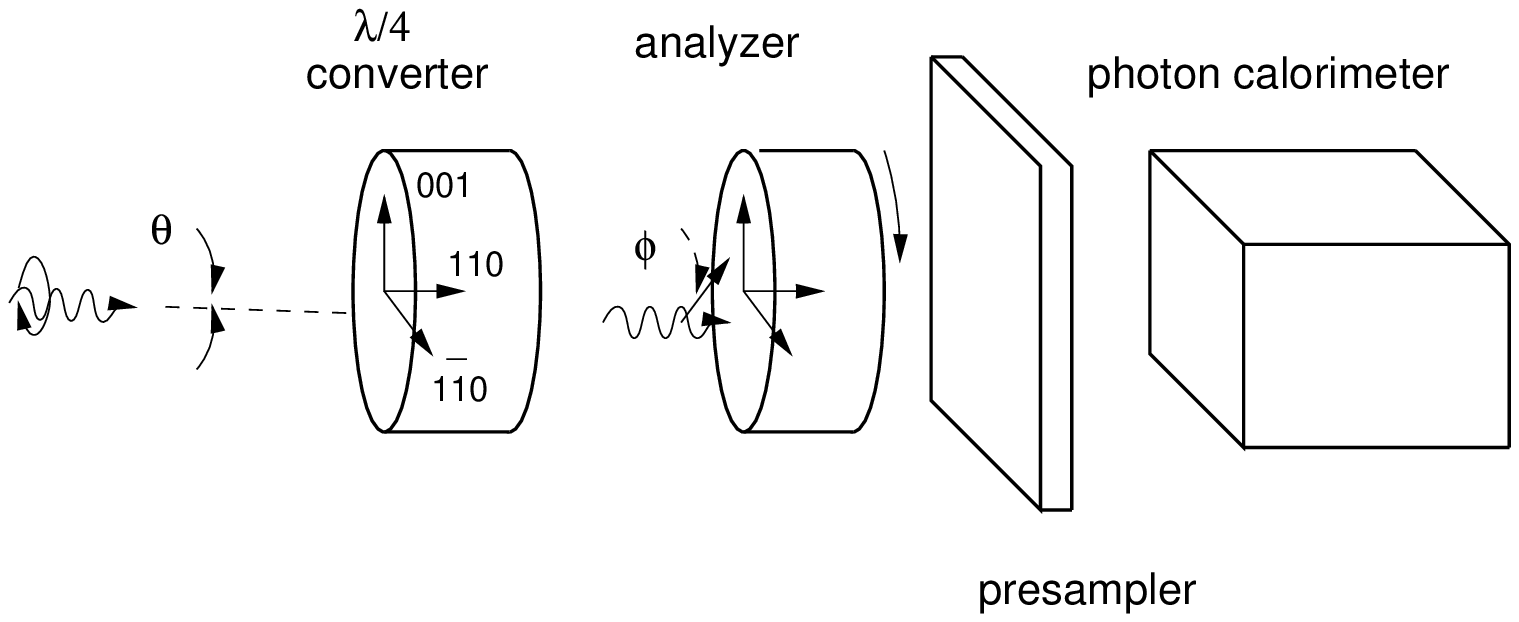, width=6cm,height=12cm,angle=-90}
\caption{\sf Modified photon detector in the ZEUS luminosity monitor
which allows the measurement of the bremsstrahlung polarization.}
\label{setup}
\efg

Both crystals would be installed on remotely controlled  goniometers,
providing three degrees of freedom. A 3~$X_0$ thick
$\lambda/4$ plate would be installed at a fixed (but adjustable) angle between
its crystallographic axes and the incident photon direction. A 1.5--2~$X_0$
thick analyzer would be placed directly behind the converter with a small angle
between one of its symmetry axes and the incident photon beam direction, and
would continuously rotate about the photon beam axis. To detect photon
interactions in the crystals the lead plate in front of the presampler has to
be removed and an additional separate readout must be installed. The detector
would tag a photon interaction in the crystals if the energy deposit in the
presampler would exceed the signal from one minimum ionizing particle.
\bfg
\includegraphics{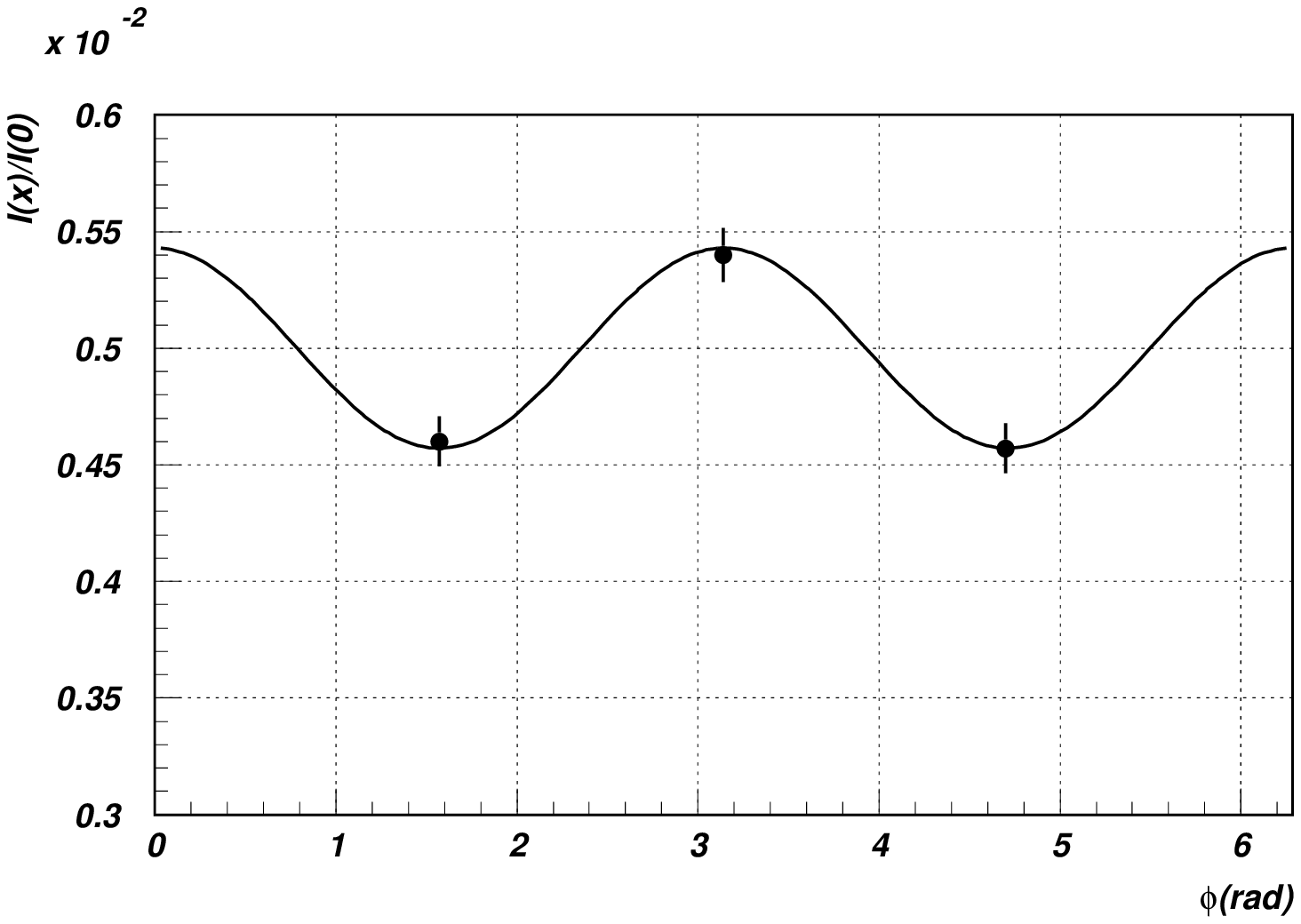}
\begin{picture}(140,75)(0,0)
\end{picture}
\caption{\sf Variation with $\phi$ of the flux of photons which reach the
photon calorimeter without interaction in the crystals (normalized
to the incident photon flux); the error bars with dots
represent a statistical accuracy achievable in a 1~min measurement at a given
$\phi$ angle setting; $P_\gamma^C=0.5$ was assumed.}
\label{data}
\efg

In Fig. \ref{data} the number of photons reaching the photon
calorimeter (with no signal in the presampler) is shown as
a function of the analyzer rotation angle $\phi$. This result
was obtained using Eqs. \ref{ceq},\ref{aeq} and assuming
$3~X_0$ converter and $1~X_0$ analyzer and
A=0.14, ${\rm A}_c=0.41$.
\section{Experimental Issues}
The transverse size of the photon beam at the photon detector
is about 8~cm horizontally and 6~cm vertically which requires
crystals with cross-sections of about $10\times10~cm^2$. The total
thickness of the crystals is limited to about 5~$X_0$ to keep
the photon beam attenuation at an acceptable level ($>$1/50)
and to avoid large non-linearities in the photon energy measurement.
There are only three types of crystals available in such
large bulk quantities: highly oriented pyrolytic graphite, and silicon
and germanium crystals. Pyrolytic graphite behaves as a single crystal
only in one dimension and was used in the Cornell and SLAC experiments.
\bfg
\includegraphics{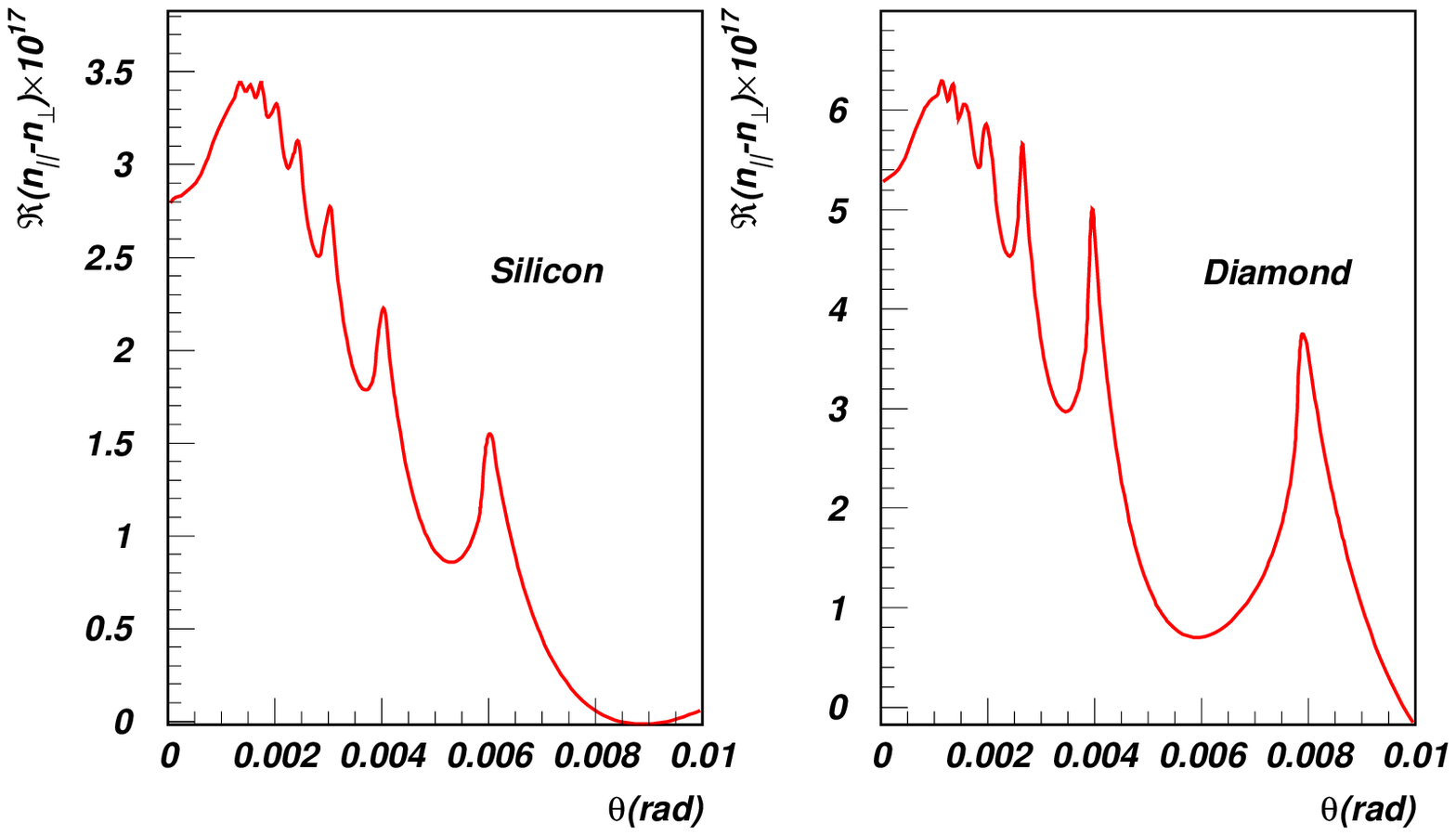}
\begin{picture}(140,75)(0,0)
\end{picture}
\caption{\sf Results of calculations of \dn\ using analytic formulae
from $[8,11]$ for silicon and diamond crystals aligned along (110) axis.}
\label{dn}
\efg

In Tab. \ref{input} the crystal parameters used in the calculation of
\dn\ for five crystal types (diamond and copper were considered
for reference) are shown, following the notation and definitions
in \cite{cabbibo3,cabbibo2}. In Fig. \ref{dn} \dn\ in
diamond and silicon crystals is shown for 27~GeV photons as
a function of the incident angle, $\theta$, measured with respect to the
(110) axis\footnote{The re-calculation of \dn\ for
a copper crystal using analytic formulae from \cite{cabbibo2,cabbibo3}
yielded results different than those given in \cite{cabbibo2}
(as was also observed in \cite{antonio}). We believe that it is due
to some numerical mistake in \cite{cabbibo2} as we found good agreement
between our calculations and results published in \cite{maisheev} for
\dn\ in silicon crystals.}. The distributions show similar behavior
-- series of narrow, small peaks on top of a broad (about 1~mrad wide) peak
at 1.5--2~mrad incident angle. Additional smearing is expected due to spread
of direction of crystal axes (`mosaic spread'), which is particularly large for
graphite crystals.

\begin{table}
\begin{center}
\caption{\sf Crystal Parameters Used in \dn\ Calculation}
\label{input}
\begin{tabular}{l|r r r}
Crystal &Lattice Constant& Screening Parameter & Thermal Factor\\
(Lattice Type)& a ($m_e^{-1}$)& b
($m_e^{-2}$)& A ($m_e^{-2}$)\\
\hline
Graphite (pyrolytic) &  869 & 3731 & 227\\
Silicon (diamond) & 1400 & 2112 & 282\\
Copper (fcc, at 77 K) & 935 & 1360 & 166\\
Germanium (diamond) & 1463 & 1316 & 286\\
Diamond & 922 & 3731 & 126\\
\end{tabular}
\end{center}
\end{table}
\begin{table}
\begin{center}
\caption{\sf Results of Calculations}
\label{results}
\begin{tabular}{l|r r}
Crystal Type & d ($X_0$) & $\theta_0$~(mrad) \\
\hline
Graphite (pyrolytic) &  $>15$ & 1.7\\
Silicon (diamond) & 3.5 & 1.5\\
Copper (fcc, at 77 K) & 1.0 & 1.2\\
Germanium (diamond) & 4.4 & 1.3\\
Diamond & 1.5 & 1.1\\
\end{tabular}
\end{center}
\end{table}
In Tab. \ref{results} results are given for the thickness of the $\lambda/4$
converter, d, at the best angle
of incidence, $\theta_0$, for 27~GeV photons (for all crystals $\theta_0$
was measured between the photon momentum and the (110) axis, except for
the graphite crystal where it was measured with respect to (002) axis).
It is clear that the required converter thickness (in radiation lengths)
decreases for heavier crystals favoring usage of silicon
or germanium crystals. The converter thickness is unfortunately
rather large, which does not leave much room for the analyzer.

The crystals would be subjected to huge synchrotron radiation dosages of
$>10^6$~Gy/year which should not pose major problems according to
recent results from SPS beam extraction experiments where $10^{20}$
protons/$cm^2$ fluences (equivalent to about $10^{11}$~Gy energy deposits)
did not change significantly the performance of silicon crystals \cite{sps}.
As synchrotron light would deposit about 300~W power in the crystals,
the crystals might require cooling (as in the SLAC experiment \cite{eisele})
to avoid large temperature-dependent effects.

The crystal alignment is not critical (of the order of $\pm50 \mu$rad) due
to the photon beam divergence ($\approx100 \mu$rad) and the crystal
mosaic spread of at least 3~mrad for pyrolytic graphite, for example.

The rotating analyzer will cause a time dependent photon absorption and
could bias the simultaneous luminosity measurement, however recent results
from CERN indicate that the change of energy absorption in crystal
is small \cite{absorption}. Additionally, if the rotation is uniform and
not too slow (i.e. faster than the updates of the integrated
luminosity every 16~s or so) this effect should average out.

Beside many experimental effects which would have to be well
controlled, as efficiency of the presampler or contribution of the
backgrounds and accidental coincidences, there are three major
parameters which have to be calibrated: converter phase shift,
$\alpha=\dn\Eg d$, and the analyzing powers $A_c(d)$ of the converter and
$A(x)$ of the analyzer. If the crystals are made of the same material,
A and ${\rm A}_c$ can be measured using standard procedures --
measurement of the product, $A A_c$, polarizing linearly an initially
unpolarized photon beam with the converter and measuring the resulting
polarization with the analyzer \cite{clarke,lep}. Then, A and ${\rm A}_c$
can be resolved according to the crystal thicknesses.

The phase shift measurement is more involved and we propose to utilize
the converter behavior for energies different than the nominal energy
when $\alpha$ should be equal to $\pi/4$. It would require the simultaneous
measurement of the
bremsstrahlung polarization at the electron beam energy and, for example,
at a half of the beam energy. For y=0.5 $P_\gamma^C=0.64P_e^L$, and A,
$A_c$ can be predicted and/or calibrated. The remaining change of the
measured polarization will be due to the change of the phase shift.
This
effect may be used for the measurement of the nominal phase shift according
to the following formula:
\beq
\cos^2\frac{\alpha}{2}=\frac{(0.64)^2}{4}
\frac{P^2_h-A_{c,h}^2}{P^2_l-A_{c,l}^2}\frac{1-A_{c,l}^2}{1-A_{c,h}^2}\, ,
\eeq
where $A_{c,h}, A_{c,l}$ are the converter analyzing powers, and
$P_h, P_l$ are the linear polarizations measured with analyzer, at
high and low photon energies, respectively. Such a calibration procedure
can be repeated at many photon energies and its precision is limited only
by the uncertainty of the analyzing powers A and $A_c$, as the energy
scale in the ZEUS luminosity monitor is precisely controlled \cite{bse95}.
Most probably the calibration data would be taken
using electron-gas bremsstrahlung which offers more stable and clean
experimental conditions, although at the cost of the data statistics.
\section{Conclusions}
Polarized high energy bremsstrahlung at HERA offers unique opportunity for
studying coherent effects in crystals. Existing detectors used in the
ZEUS experiment for the luminosity measurement, after some modifications,
would allow for detailed studies of these effects in a 5--30~GeV photon
energy range. The calibration procedures of the relevant
crystal parameters, in particular the calibration procedure of crystal
phase shift proposed in this paper, could be done with experimental data
rather than with theoretical models. This opens
opportunity of the novel technique of the measurement of the the electron
beam longitudinal polarization using crystals. Although this is a very
demanding task in the difficult HERA environment (huge beam collision rates
and severe background conditions) we believe that polarization measurement
with 2\% precision is feasible.
\section*{Acknowledgment}
The author would like to thank Vladimir Maisheev for pointing out the
numerical problems with the published data on \dn\ and for useful
comments and explanations. I thank very much Antonio Di Domenico for
early discussions on this subject and for providing me with useful information.
I am very grateful to Mark Lomperski for critical reading of the manuscript. I
wish
to thank very much the Workshop organizers for inviting me to \AA rhus
and supporting my stay there.

\end{document}